\documentclass[conference]{IEEEtran}
\IEEEoverridecommandlockouts
\usepackage{cite}
\usepackage{amsmath,amssymb,amsfonts}
\usepackage{algorithmic}
\usepackage{graphicx}
\usepackage{textcomp}
\usepackage{xcolor}
\usepackage{tikz}
\def\BibTeX{{\rm B\kern-.05em{\sc i\kern-.025em b}\kern-.08em
    T\kern-.1667em\lower.7ex\hbox{E}\kern-.125emX}}
\begin{document}

\title{Polar Codes with Local-Global Decoding\\
}

\author{\IEEEauthorblockN{Ziyuan Zhu}
\IEEEauthorblockA{
University of California, San Diego \\
ziz050@ucsd.edu}
\and
\IEEEauthorblockN{Wei Wu}
\IEEEauthorblockA{
University of California, San Diego \\
wew128@ucsd.edu}
\and
\IEEEauthorblockN{Paul H. Siegel}
\IEEEauthorblockA{
University of California, San Diego \\
psiegel@ucsd.edu}

}

\maketitle

\begin{abstract}
In this paper, we investigate a coupled polar code architecture that supports both local and global decoding. This local-global construction is motivated by practical applications in data storage and transmission where reduced-latency recovery of sub-blocks of the coded information is required. Local decoding allows random access to sub-blocks of the full code block. When local decoding performance is insufficient, global decoding provides improved data reliability. The coupling scheme incorporates a systematic outer polar code and a partitioned mapping of the outer codeword to  semipolarized bit-channels of the inner polar codes. Error rate simulation results are presented for 2 and 4 sub-blocks. Design issues affecting the trade-off between local and global decoding performance are also discussed.

\end{abstract}
\section{Introduction}
Polar codes, proposed by Erdal Ar\i kan in 2009, provide a deterministic coding scheme that provably achieves the Shannon capacity of any  symmetric, binary-input  discrete  memoryless channel under successive cancellation (SC) decoding\cite{Arikan2009}. They have attracted enormous interest and have been incorporated into the 5G New Radio wireless standard as the control channel coding scheme. Belief propagation (BP) decoding of polar codes, which provides soft decoder outputs with relatively low complexity, was suggested by Ar\i kan in\cite{ArikanBP} and has since been widely investigated; see, e.g., \cite{Niu2014}. 

Systematic encoding of polar codes was introduced by Ar\i kan in \cite{Arikan2011} for use in scenarios where it is desirable for the encoded information to appear explicitly in the codeword. Moreover, he showed empirically that SC decoding with re-encoding offered a superior bit error rate performance than non-systematic encoding. This performance improvement has also been observed under BP decoding. 

In\cite{Guo2014}, Guo et al. proposed enhanced BP decoding of polar codes through concatenation of an outer code that protects bit-channels of intermediate channel quality, referred to as semipolarized bit-channels. To illustrate the idea, they considered an outer LDPC code and an outer convolutional code. Elekelesh et al.\cite{Elk2017} extended this idea and introduced an augmented polar code construction using an  auxiliary outer polar code to protect semipolarized bit-channels. They also proposed a flexible-length polar code construction that couples two polar codes of different lengths through such an auxiliary outer polar code. 

In \cite{RamCas2018}, Ram and Cassuto proposed a coupling architecture for LDPC codes that supports two levels of decoding: local decoding of sub-blocks for use in good channel conditions and global decoding of the coupled codewords for use under adverse channel conditions. 
In this paper, we propose a modification of the polar code coupling architecture in~\cite{Elk2017} that supports such local and global decoding. 

The paper is organized as follows. Section II briefly reviews the background concepts of channel polarization, BP decoding of polar codes,  and systematic polar codes. In Section III, the local-global decoding architecture for polar codes is introduced. Section IV provides bit error rate (BER) and frame error rate (FER) simulation results demonstrating the performance of the local-global decoding architecture, as well as a discussion of design issues that affect the trade-off between local and global decoding performance. Section V concludes the paper.

\section{Background} 

\subsection{Channel Polarization}
The construction of polar codes and their capacity-achieving properties under SC decoding are based upon channel polarization.  
Channel polarization includes operations of channel combining and channel splitting: $N$ independent copies of a channel $W$ are combined in a recursive manner  into a vector channel $W_N$, which is then split into $N$ synthesized channels ${W_N^{(i)}, \;1\leq{i}\leq{N}}$, referred to as bit-channels.
Let $G_N=F^{\bigotimes n}$ be the $N\times N$ matrix that is the $n^{th}$ Kronecker power of $F=\left[\begin{array}{cc}1 & 0 \\1 & 1 \\ \end{array} \right]$, where $n=\log_2 N$.
Given ${\bf u} \in \{0,1\}^N$, the vector channel is characterized by $W_N({\bf y}|{\bf u})=W^N({\bf y}|{\bf x} )$, where $W^N$ denotes the product channel corresponding to $N$ independent uses of channel $W$, and  ${\bf x}={\bf u}G_N$.   The bit-channels are defined by  $W_N^{(i)}({\bf y}, u_1^{i-1}|u_i)=\sum_{u_{i+1}^{N}} \frac{1}{2^{N-1}} W_N({\bf y}|{\bf u})$, for $1\leq{i}\leq{N}$. As $N\rightarrow \infty$, the channel polarization theorem states that the Bhattacharyya parameter $Z(W_N^{(i)})$ converges to either 0 or 1. A polar code of rate $R=K/N$ uses the $K$ bit-channels with the lowest $Z(W_N^{(i)})$ for the information bits, and the remaining bits are frozen to value zero. We use ${\mathcal A}$ to denote the information indices, and ${\mathcal F=A^c}$ to denote the  frozen indices.

\tikzstyle{var} = [circle, fill, minimum size=3pt,inner sep=0pt, outer sep=0pt]
\tikzstyle{add} = [circle, draw, minimum size=2pt, inner sep=0pt]
\tikzstyle{connect} = [circle, draw, minimum size=2pt, inner sep=-1pt]
\tikzstyle{every node}=[font=\footnotesize]
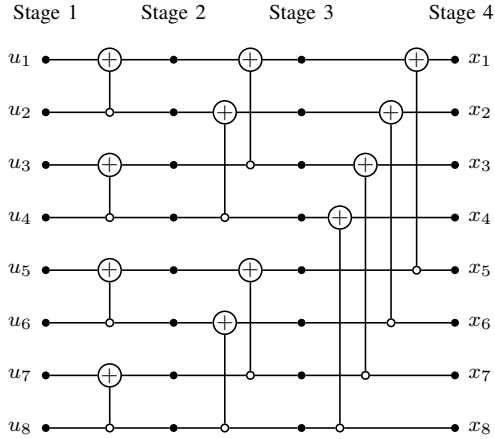
\begin{figure}
    \centering
\begin{tikzpicture}[yscale=0.7, xscale=0.85, node distance=0.2cm, auto, semithick]
	\foreach \y in {1,...,8}
        \node[var] (0-\y) [label=left:$u_{\y}$] at (0,-\y) {};
	\foreach \y in {1,...,8}
        \node[var] (6-\y) [label=right:$x_{\y}$] at (6.4,-\y) {};
	\foreach \y in {1,3,5,7}
		\node[add] (1-\y) [] at (1,-\y) {$+$};       
    \foreach \y in {2,4,6,8}
		\node[connect] (1-\y) [] at (1,-\y) {};   	
	\foreach \y in {1,...,8}
		\path (0-\y) edge[-] (1-\y);
	\foreach \y / \x in {1/2, 3/4, 5/6, 7/8}
		\path (1-\y) edge[-] (1-\x);
  
	\foreach \y in {1,...,8}
        \node[var] (2-\y) [] at (2,-\y) {};	
    \foreach \y in {1,...,8}
		\path (1-\y) edge[-] (2-\y);
	\foreach \y in {2, 6}
		\node[add] (3-\y) [] at (2.8,-\y) {$+$};
	\foreach \y in {1, 5}
		\node[add] (3-\y) [] at (3.2,-\y) {$+$};
	\foreach \y in {4, 8}
		\node[connect] (3-\y) [] at (2.8,-\y) {};
	\foreach \y in {3, 7}
		\node[connect] (3-\y) [] at (3.2,-\y) {};
	\foreach \y in {1,...,8}
        \node[var] (4-\y) [] at (4,-\y) {};		
	\node[add] (5-1) [] at (5.8,-1) {$+$};		
	\node[add] (5-2) [] at (5.4,-2) {$+$};	
	\node[add] (5-3) [] at (5.0,-3) {$+$};	
	\node[add] (5-4) [] at (4.6,-4) {$+$};	
	\node[connect] (5-5) [] at (5.8,-5) {};		
	\node[connect] (5-6) [] at (5.4,-6) {};	
	\node[connect] (5-7) [] at (5.0,-7) {};	
	\node[connect] (5-8) [] at (4.6,-8) {};

	\foreach \y in {1,...,8}
		\path (2-\y) edge[-] (3-\y);
	\foreach \y in {1,...,8}
		\path (3-\y) edge[-] (4-\y);
	\foreach \y in {1,...,8}
		\path (4-\y) edge[-] (5-\y);	
	\foreach \y in {1,...,8}
		\path (5-\y) edge[-] (6-\y);
	\path (3-3) edge[-] (3-1);
	\path (3-4) edge[-] (3-2);
	\path (3-7) edge[-] (3-5);
	\path (3-8) edge[-] (3-6);
	\path (5-5) edge[-] (5-1);
	\path (5-6) edge[-] (5-2);
	\path (5-7) edge[-] (5-3);
	\path (5-8) edge[-] (5-4);
	
	\node [above] at (0,-0.5) {Stage 1};
	\node [above] at (2,-0.5) {Stage 2};
	\node [above] at (4,-0.5) {Stage 3};
	\node [above] at (6.5,-0.5) {Stage 4};
\end{tikzpicture}
    \caption{Encoder-based factor graph for length $N=8$ polar code.}
    \label{encoder}
\end{figure}

\subsection{BP Decoding}
BP decoding of polar codes is an iterative message passing algorithm that operates on the sparse factor graph derived from the encoder structure,  illustrated in Fig. \ref{encoder} for $N=8$.
The factor graph is composed of basic substructures corresponding to the combining operation represented by the matrix $F$, as shown in Fig.~\ref{substructure}. Two types of log-likelihood ratio (LLR) messages are generated and passed bidirectionally:  $L$-messages that are passed from right-to-left and $R$-messages that are passed from left-to-right. The $L$-messages at the rightmost nodes represent the channel LLRs. The $R$-messages at the leftmost modes are assigned $0$ or $\infty$ if they are in $\mathcal{A}$ or $\mathcal{F}$, respectively. All other $L$-messages and $R$-messages are initialized to $0$. Messages are propagated from right-to-left and then from left-to-right, with the following message update rules~\cite{YZhang2014} applied at each substructure 

\tikzstyle{var} = [circle, fill, minimum size=3pt,inner sep=0pt, outer sep=0pt]
\tikzstyle{add} = [circle, draw, minimum size=2pt, inner sep=0pt]
\tikzstyle{connect} = [circle, draw, minimum size=2pt, inner sep=-1pt]
\tikzstyle{every node}=[font=\footnotesize]
\begin{figure}
    \centering
\begin{tikzpicture}[yscale=1.5, xscale=2, node distance=0.2cm, auto, semithick]
	\foreach \y in {0,1}
        \node[var] (0-\y) at (0,-\y) {};
	\foreach \y in {0}
		\node[add] (1-\y) [] at (1,-\y) {$+$};
    \foreach \y in {1}
		\node[connect] (1-\y) [] at (1,-\y) {};
	\foreach \y in {0,1}
		\path (0-\y) edge[-] (1-\y);
	\foreach \y / \x in {1/0}
		\path (1-\y) edge[-] (1-\x);
	\foreach \y in {0,1}
        \node[var] (2-\y) [] at (2,-\y) {};	
    \foreach \y in {0,1}
		\path (1-\y) edge[-] (2-\y);
		
\draw [->, blue] (0.9,0.1) -- (0.1,0.1);
\draw [->, blue] (1.9,0.1) -- (1.1,0.1);
\draw [->, blue] (0.9,-0.9) -- (0.1,-0.9);
\draw [->, blue] (1.9,-0.9) -- (1.1,-0.9);
\node [above,blue] at (0.5,0.1) {$L_{out,1}$};
\node [above,blue] at (1.5,0.1) {$L_{in,1}$};
\node [above,blue] at (0.5,-0.9) {$L_{out,2}$};
\node [above,blue] at (1.5,-0.9) {$L_{in,2}$};

\draw [->, red] (0.1,-0.1) -- (0.9,-0.1);
\draw [->, red] (1.1,-0.1) -- (1.9,-0.1);
\draw [->, red] (0.1,-1.1) -- (0.9,-1.1);
\draw [->, red] (1.1,-1.1) -- (1.9,-1.1);

\node [below,red] at (0.5,-0.1) {$R_{in,1}$};
\node [below,red] at (1.5,-0.1) {$R_{out,1}$};
\node [below,red] at (0.5,-1.1) {$R_{in,2}$};
\node [below,red] at (1.5,-1.1) {$R_{out,2}$};
\end{tikzpicture}
    \caption{Basic substructure.}
    \label{substructure}
\end{figure}
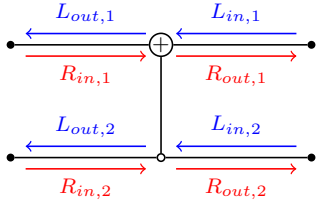






\vspace{-1em}
\begin{equation}
\begin{split}
    L_{out,1} = L_{in,1} \boxplus \left( L_{in,2}+R_{in,2} \right)\\
    R_{out,1} = R_{in,1} \boxplus \left( L_{in,2}+R_{in,2} \right)\\
    L_{out,2} = \left( R_{in,1} \boxplus L_{in,1} \right) +L_{in,2}\\
    R_{out,2} = \left( R_{in,1} \boxplus L_{in,1} \right) +R_{in,2}
\end{split}
\end{equation}
where the  box-plus  operator $\boxplus$ is defined as $a \boxplus b=\ln{(\frac{1+e^{a+b}}{e^{a}+e^{b}})}$.
(Other operators, such as min-sum, are sometimes used instead of box-plus.) 
Decoder output  $\hat{\bf u}$ (resp. $\hat{\bf x}$) is obtained after the final iteration by summing and thresholding the leftmost (resp. rightmost)  $L$-messages  and $R$-messages.
\subsection{Systematic Polar Codes}
In systematic polar codes~\cite{Arikan2011}, the information bits appear in specified locations in the polar codeword.  
Let ${\bf v} \in \{0,1\}^K$ represent the information bits to be encoded. The systematic encoder solves the equation
\begin{equation}{\bf x}={\bf u} F^{\bigotimes n}
\label{sys_encoder}
\end{equation}
subject to ${\bf u}_{\mathcal{F}}={\bf 0}$ and 
${\bf x}_{\mathcal{A}} = {\bf v}$.
Vangala et al.\cite{Vangala2016} present three efficient algorithms to solve for ${\bf u}_{\mathcal{A}}$ and ${\bf x}_{\mathcal{F}}$. In this paper, we use the encoder denoted ${\bf EncoderA}$.

Systematic polar codes use the same decoding algorithms as conventional polar codes, e.g., SC or BP decoding, to generate an estimate ${\bf \hat{u}}$ of the vector ${\bf u}$. To recover an estimate ${\bf \hat{v}}$ of the information bits, we compute ${\bf \hat{x}} = {\bf \hat{u}}G_N$ and set 
${\bf \hat{v}}={\bf \hat{x}}_{\mathcal{A}}$.

\section{Local-Global Polar Decoding}

Finite-length polar codes exhibit incomplete channel polarization. As mentioned above, the design of a rate $R=\frac{K}{N}$ polar code involves selecting the information indices $\mathcal{A}$ with the smallest Bhattacharyya parameters and the complementary set $\mathcal{F}{=}\mathcal{A}^c$ of frozen indices. In~\cite{Guo2014} and~\cite{Elk2017}, a finer grouping of bit-channel indices was proposed, in which, for thresholds $0 < \delta_1 \leq \delta_2 <1$, the \textit{good} bit-channel indices satisfy $Z(W_N^{(i)}) \leq \delta_1$, the \textit{frozen} bit-channel indices satisfy  $Z(W_N^{(i)}) > \delta_2$, and the \textit{semipolarized} bit-channel indices satisfy $\delta_1 < Z(W_N^{(i)}) \leq \delta_2$. An auxiliary outer code, such as an LDPC block code~\cite{Guo2014} or a polar code~\cite{Elk2017}  is then used to protect the vulnerable bits encoded through the semipolarized bit-channels, with the (interleaved) outer codeword   providing the input bits to the semipolarized bit-channels. A natural enhancement of BP decoding, applied to a combined factor graph representing the concatenation of  the inner and outer codes, can be used to decode the information bits of the inner code and outer code. We remark that the Bhattacharyya parameter, though usually a measure for channel reliability under SC decoding, also provides a useful measure for classifying bit-channels under BP decoding\cite{Guo2014}.

The augmented polar code construction in~\cite{Elk2017}  was extended to a flexible-length polar code construction using a coupled code architecture, in which the semipolarized bit-channels of two (or more) inner polar codes are protected by an interleaved  auxiliary outer polar code. Again, an enhanced BP decoding algorithm can be used to decode the information bits of the coupled inner codes and the auxiliary code. 

Two modifications of the coupled construction in~\cite{Elk2017} provide an architecture suitable for local-global decoding. First, the auxiliary outer code is required to be systematic. Second, the interleaver maps specified subsets of the information bits embedded in the auxiliary codeword
to the semipolarized bit-channels of each of the inner codes.  We now describe in more detail the encoder and decoder for this local-global architecture.   

\subsection{Encoder}

\begin{figure}[htbp]
\centerline{\includegraphics[width=8cm,height=5cm]{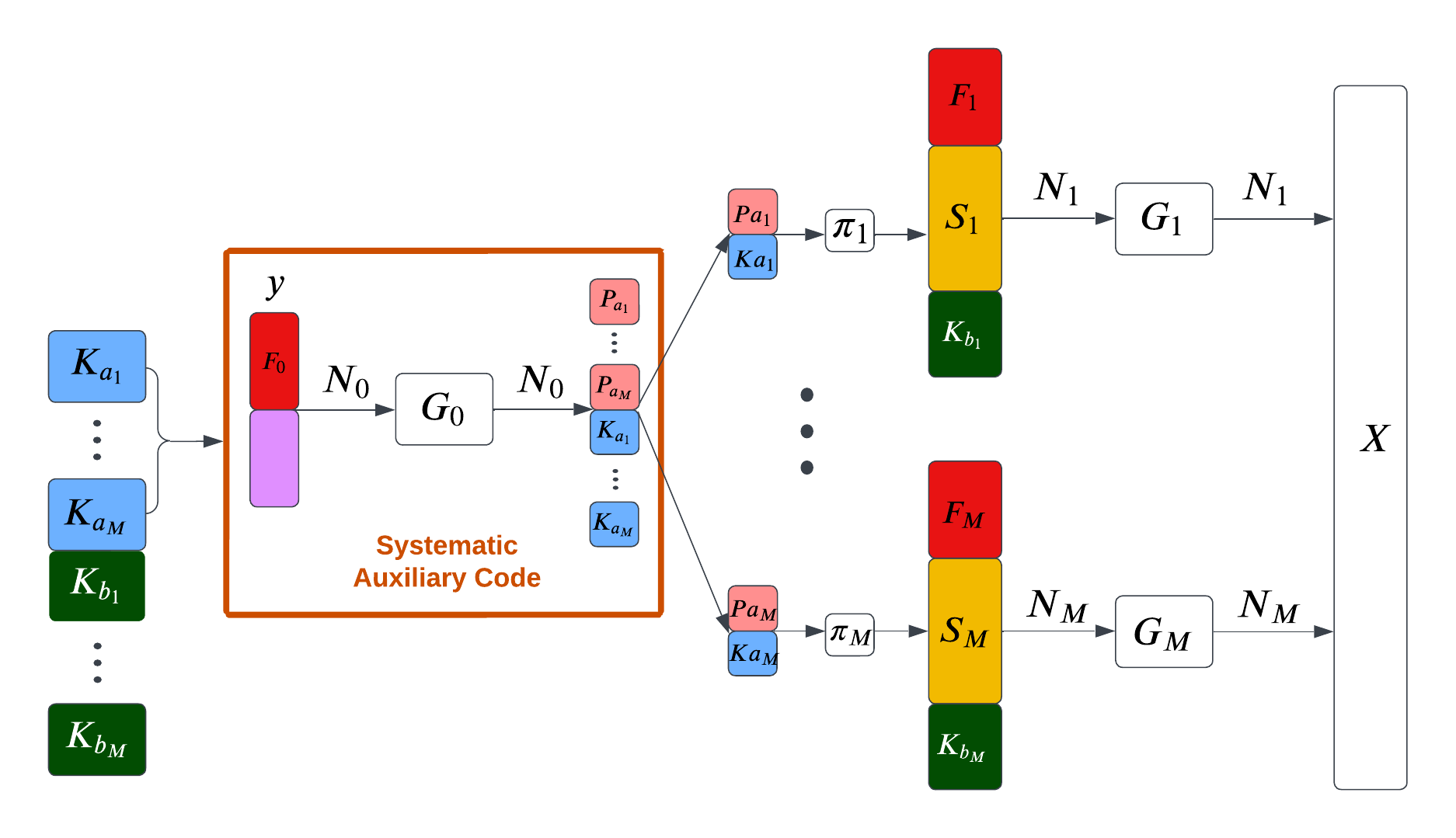}}
\caption{Encoder for local-global polar code.}
\label{local global encoder}
\end{figure}

Let $[K_a,K_b] = [K_{a_1},...,K_{a_M},K_{b_1},...,K_{b_M}]$ denote the information bits to be encoded. For purposes of illustration, we assume the sets $K_{a_i}$, $i=1, \ldots, M$ are of equal size, and similarly for the sets $K_{b_i}$, $i=1, \ldots, M$. The inputs to the systematic outer polar encoder are information bits $K_a$   and frozen bits $F_0$, and the length-$N_0$ output codeword is $[P_a,K_a]$. This codeword contains the information bits $K_a$, in known positions, and the parity bits $P_a$. Dividing the parity bits $P_a$ into $M$ equal-size subsets, the interleaver maps $[P_{a_i},K_{a_i}]$ to the semipolarized bit-channels $S_i$ of the $i^{th}$ inner code. (We also assume in this illustration that the inner codes have equal lengths.) The goal of the interleaving is to decorrelate the LLRs used in the BP decoding as much as possible in the early decoder iterations. 

The  information bits $K_{b_i}$ and additional frozen bits $F_i$ provide the input to the good bit-channels and frozen bit-channels of the $i^{th}$ inner code, respectively. The codewords of the $M$ inner codes are concatenated to form a length-$N$ codeword that is transmitted over the channel. 

A schematic of the proposed encoder for the local-global polar code is shown in Fig. \ref{local global encoder}.

The relevant code rates are as follows, where, as a shorthand notation, we use the name of a subset to represent its  cardinality. 
\begin{itemize}
    \item Combined code rate: $R_{total} = \frac{K_a+K_b}{N}$
    \item Systematic outer polar code rate: $R_{outer} = \frac{K_a}{N_0}$
    \item  $i^{th}$  inner polar code rate: $R_{inner,i} = \frac{K_{b_i}+S_i}{N_i}$
    \item $i^{th}$ subblock rate: $R_{subblock,i}=\frac{K_{b_i}+K_{a_i}}{N_i}$
\end{itemize}


\subsection{Decoder}
The architecture proposed in Fig.~\ref{local global encoder} permits separate local decoding of the inner codes, with the option of invoking global decoding of the coupled codes when local decoding does not provide satisfactory performance. Note that this architecture also retains flexibility in the choice of inner code lengths $N_i$, if that is desired. 

\subsubsection{Local decoding}
Any soft decoding scheme for polar codes can be used as a local decoding method. The estimated bits on the semipolarized bit-channels must be deinterleaved to recover the information bits $K_a$. In our simulations, we use BP decoding with early stopping as the local decoding method.

\subsubsection{Global decoding}
\begin{figure}[htbp]
\centerline{\includegraphics[width=7cm,height=6cm]{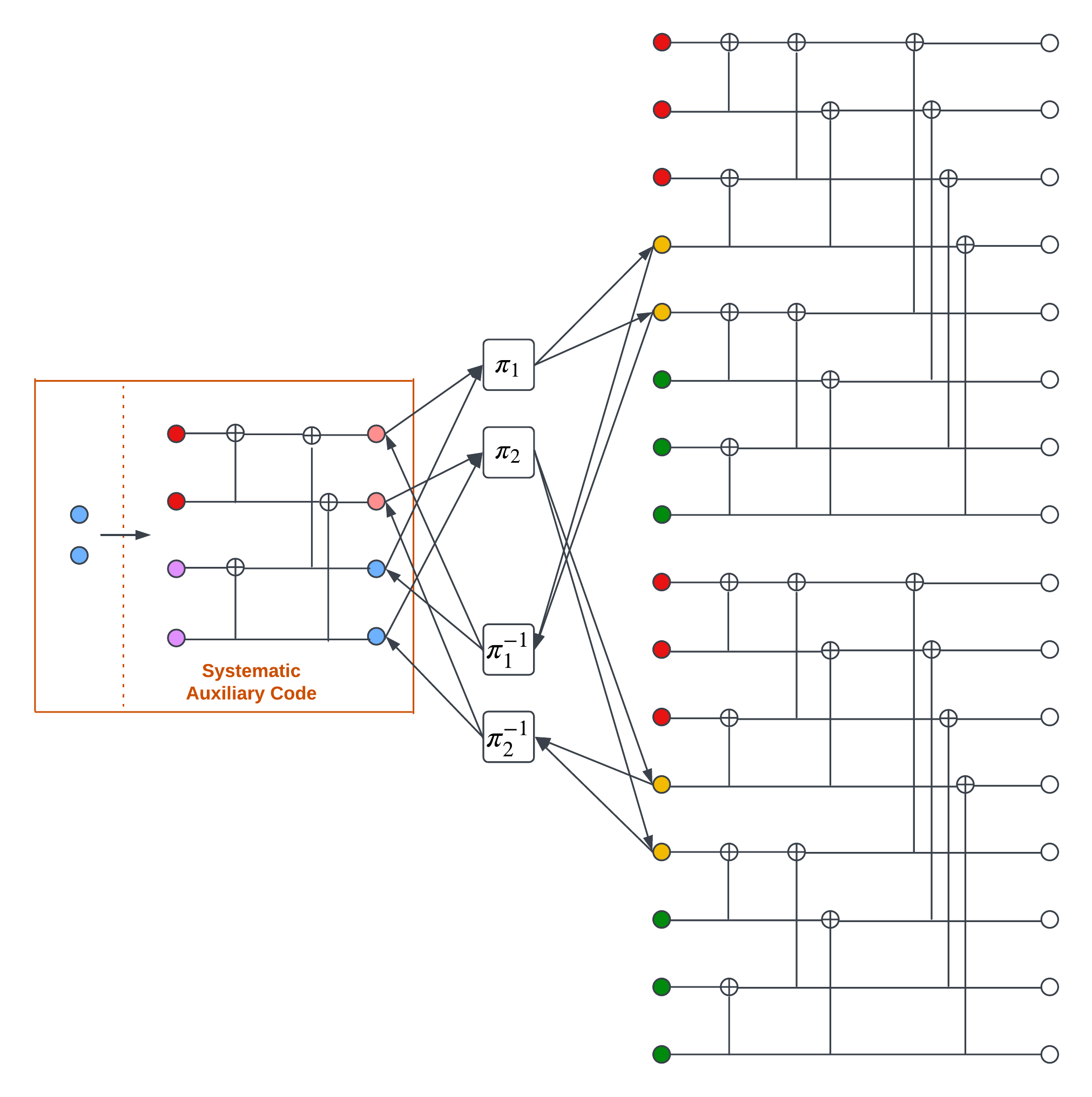}}
\caption{Factor graph for local-global polar decoding ($M = 2$).}
\label{factor graph for local global}
\end{figure}

The global decoding is carried out using BP decoding on the combined factor graph of the inner codes and outer code. Fig.~\ref{factor graph for local global} illustrates a factor graph for $M=2$ inner codes. 

For $i=1, \ldots, M$, denote the left and right LLR-messages of the $i^{th}$ inner code by $L_i$ and $R_i$, respectively, and the left and right LLR-messages of the systematic outer polar code by $L_0$ and $R_0$, respectively. An enhanced BP decoding procedure similar to that proposed in \cite{Elk2017} is used, but modified to reflect the systematic outer polar code and the partitioned interleaver. When a maximum number of iterations is specified, the decoding algorithm proceeds as follows.
\begin{enumerate}
\item
The inner BP polar decoder receives the channel LLR vector $L_{ch}$. The initial $R_i$-messages propagate from left to right; then the $L_i$-messages propagate from right to left until the leftmost stage 1 is reached.
\item
LLR-messages $L_i$ at stage 1 are passed through the corresponding  deinterleaver, and the output is passed to the BP polar decoder of the outer code. 
\item
The BP decoder of the outer polar code performs one BP iteration, with  $L_0$-messages propagating from right to left, and $R_0$-messages propagating from left to right.
\item
Next, the LLR-messages at the rightmost stage of the outer BP decoder are passed through the partitioned interleaver to the BP decoder of the inner codes.  This constitutes one global iteration. The process repeats until a maximum number of iterations is reached. 
\item
The LLRs are used to estimate the  information bits $[\hat{K}_a, \hat{K}_b]$ as follows.
\begin{itemize}
\item[(a)] The LLRs used to estimate the message $y$ at the input of the systematic outer polar code (as shown in Fig.~\ref{local global encoder}) are obtained by adding the left and right LLR-message $L_0$ and $R_0$ at stage 1 of the outer code factor graph. The estimate $\hat{y}$ is then re-encoded to obtain an estimate of the outer codeword, from which we extract the estimate of the information bits $\hat{K}_a$.

\item[(b)] The LLRs used to  estimated the message at the input of the $i^{th}$ inner polar code are obtained by adding the left and right LLR-messages $L_i$ and $R_i$ at stage 1 of the corresponding inner code factor graph. From this estimate, we extract the estimate of the information bits $\hat{K}_{b_i}$.
\end{itemize}
\end{enumerate}

In order to reduce decoding time, we introduce early stopping conditions during decoding.
For each of the early stopping conditions, we use the G-matrix criterion proposed in~\cite{EarlyStop}. 


With early stopping, steps 2) through 4) of the decoding procedure are modified as follows.
\begin{enumerate}
    \item[2*)]
    After step 2), the early stopping conditions are checked for each of the inner polar codes.
    \item[3*)]
    After step 3), the early stopping condition is checked for the outer systematic polar code.
    \item[4*)]
    After step 4), when a global iteration has completed, the results of the early stopping checks for all inner and outer codes
    are used to determine if decoding can be terminated. If so, skip to step 5). If not, go back to step 2).
    
    
    
\end{enumerate}

\begin{figure}[htbp]
\centerline{\includegraphics[width=9cm]{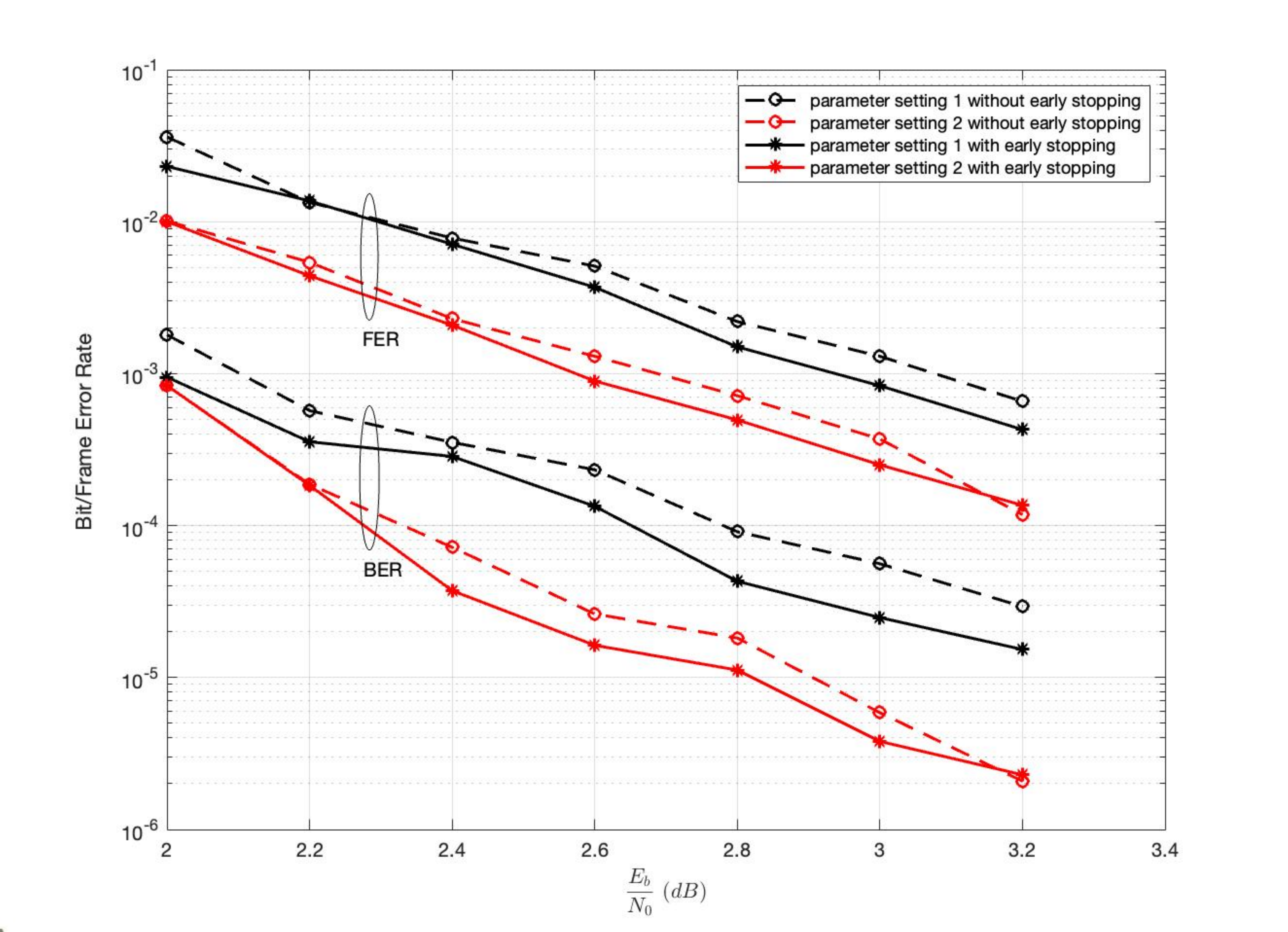}}
\caption{Global decoding with/without early stopping.}
\label{early stop result}
\end{figure}

Simulation results for global decoding with early stopping are shown in Fig.~\ref{early stop result}. Code parameter settings 1 and 2, shown in  Table~\uppercase\expandafter{\romannumeral1}, were used. Surprisingly, with early stopping, BER and FER  are both improved, in contrast to what has been observed when early stopping is used with conventional BP decoding of a polar code~\cite{EarlyStop}.  

\begin{table}[htbp]
\caption{System parameter settings.}
\begin{center}
\scalebox{1.2}{ 
\begin{tabular}{|c|c|c|c|}
\hline
\textbf{Parameters}&\textbf{Setting 1}&\textbf{Setting 2}&\textbf{Setting 3}\\
\hline
$R_{total}$ & 0.5 & 0.5 & 0.5\\
$R_{outer}$ & 0.5 & 0.5 & 0.5\\
$R_{inner}$ & 0.53125 & 0.5625 & 0.53125\\
$M$ & 2 & 4 & 4\\
$K_a$ & 64 & 256 & 128\\
$K_b$ & 960 & 1792 & 1920\\
$S_i, i\geq 1$ & 64 & 128 & 64\\
$N_0$ & 128 & 512 & 256\\
$N_i, i\geq 1$ & 1024 & 1024 & 1024\\
$Max\ iteration$ & 200 & 200 & 200\\
$Early\ stop$ & \checkmark & \checkmark & \checkmark\\
\hline
\end{tabular}
}
\label{parameter setting}
\end{center}
\end{table}

\begin{figure}[htbp]
\centerline{\includegraphics[width=9cm]{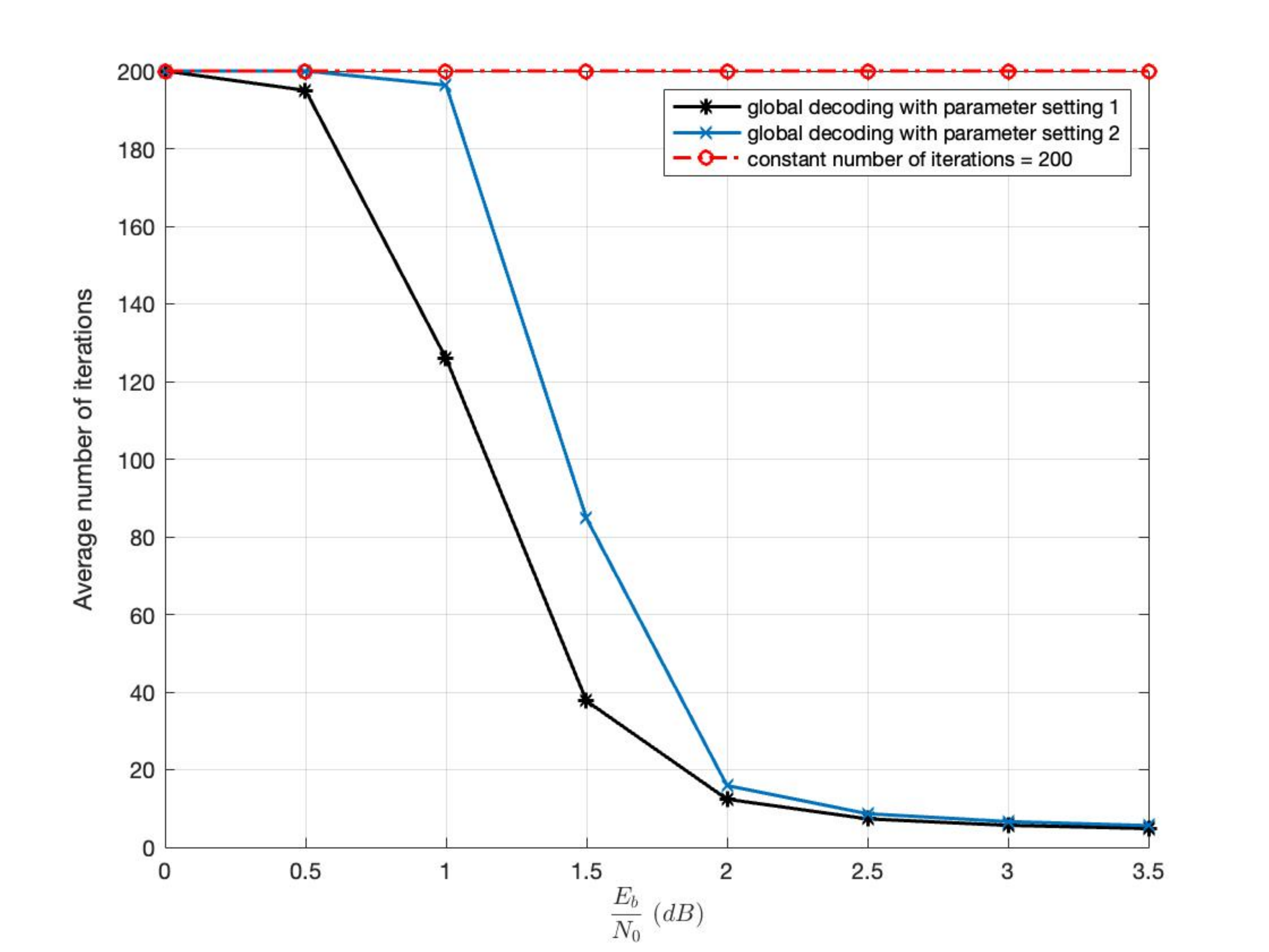}}
\caption{Average number of iterations for parameter settings 1 and 2.}
\label{early stop ave iteration}
\end{figure}

\subsection{Local-Global Decoding Simulation Results}
\begin{figure}[htbp]
\centerline{\includegraphics[width=9cm]{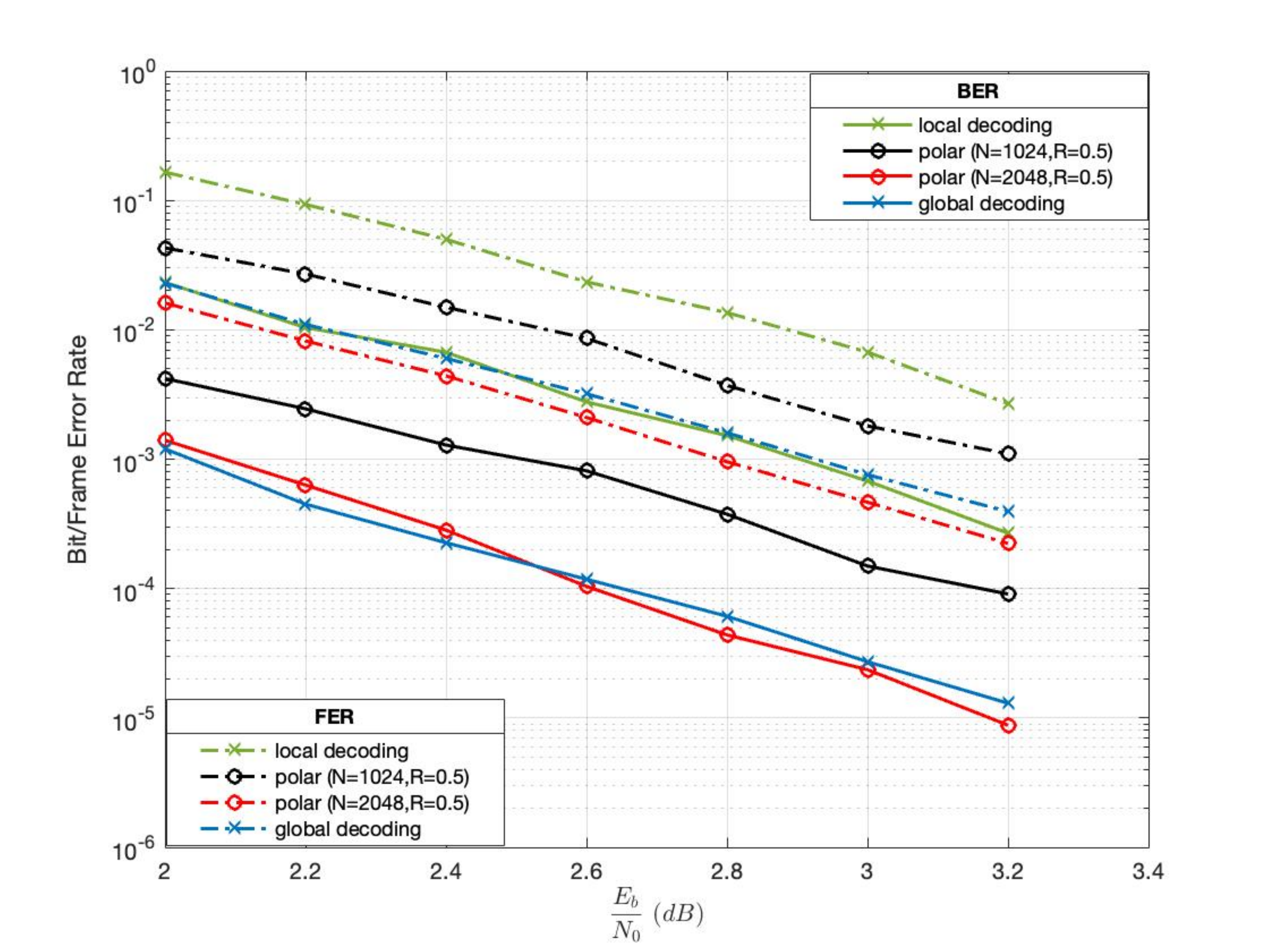}}
\caption{Local-global decoding with 2 subblocks.}
\label{2parts result}
\end{figure}

\begin{figure}[htbp]
\centerline{\includegraphics[width=9cm]{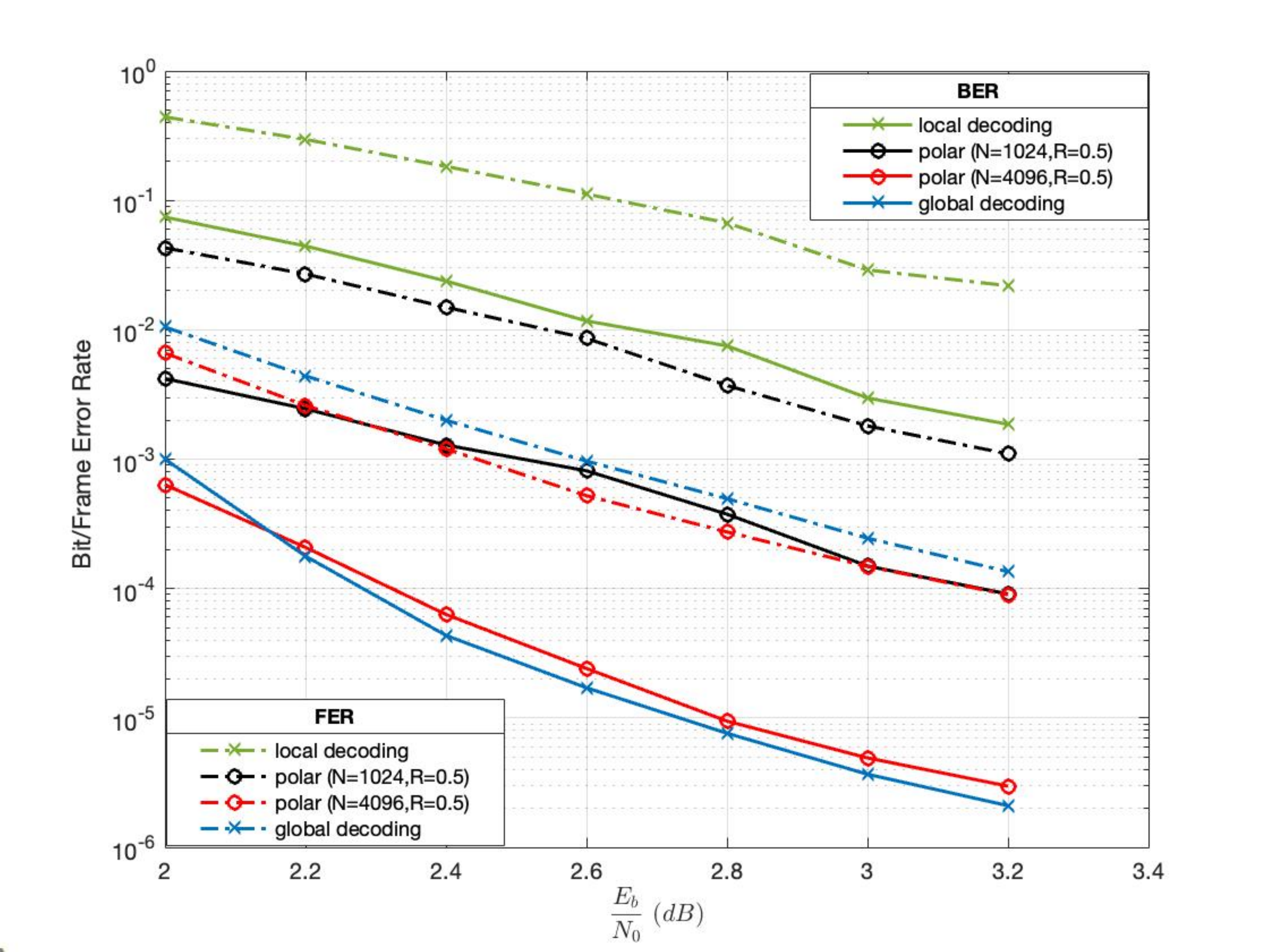}}
\caption{Local-global decoding with 4 subblocks.}
\label{4parts result}
\end{figure}

 BER and FER results for local-global decoding with 2 and 4 subblocks, each corresponding to an inner code of length 1024, are shown in Fig.~\ref{2parts result} and Fig.~\ref{4parts result}, respectively. Each figure shows the performance of a typical subblock under local decoding, along with that of a conventional polar code of length 1024 under BP decoding. Also shown is the performance of the full codeword under global decoding, along with the performance of a conventional polar code of length equal to that of the full codeword under BP decoding. Simulation results (not shown here) confirmed that the BERs and FERs of the subblocks under local and global decoding are essentially identical and that subblock failures appear to be independent.
 
 The simulations assume an AWGN channel with BPSK modulation, and the bit-channel ordering is based on Bhattacharyya bounds\cite{Arikan2009} designed at an $E_b/N_0$ of 0 dB. Early stopping is applied to both local and global decoding. 
 
Parameter setting 1 of Table~\uppercase\expandafter{\romannumeral1} was used for the 2-subblock construction, while parameter setting 2 was used for the 4-subblock construction.  

 We see that, in both cases,  global decoding significantly improves the decoding performance compared to local decoding. Global decoding also provides a BER comparable to that of the  conventional polar code of the same length and rate. In the case of 4 subblocks, global decoding actually shows a 0.1dB gain at a BER of $10^{-5}$. 
However, the benefits obtained through global decoding are achieved at the expense of  local decoding performance, with a  trade-off that can be adjusted by modifying code parameters such as $R_{inner}$, $R_{outer}$ and $R_{total}$. 


\begin{figure}[htbp]
\centerline{\includegraphics[width=9cm]{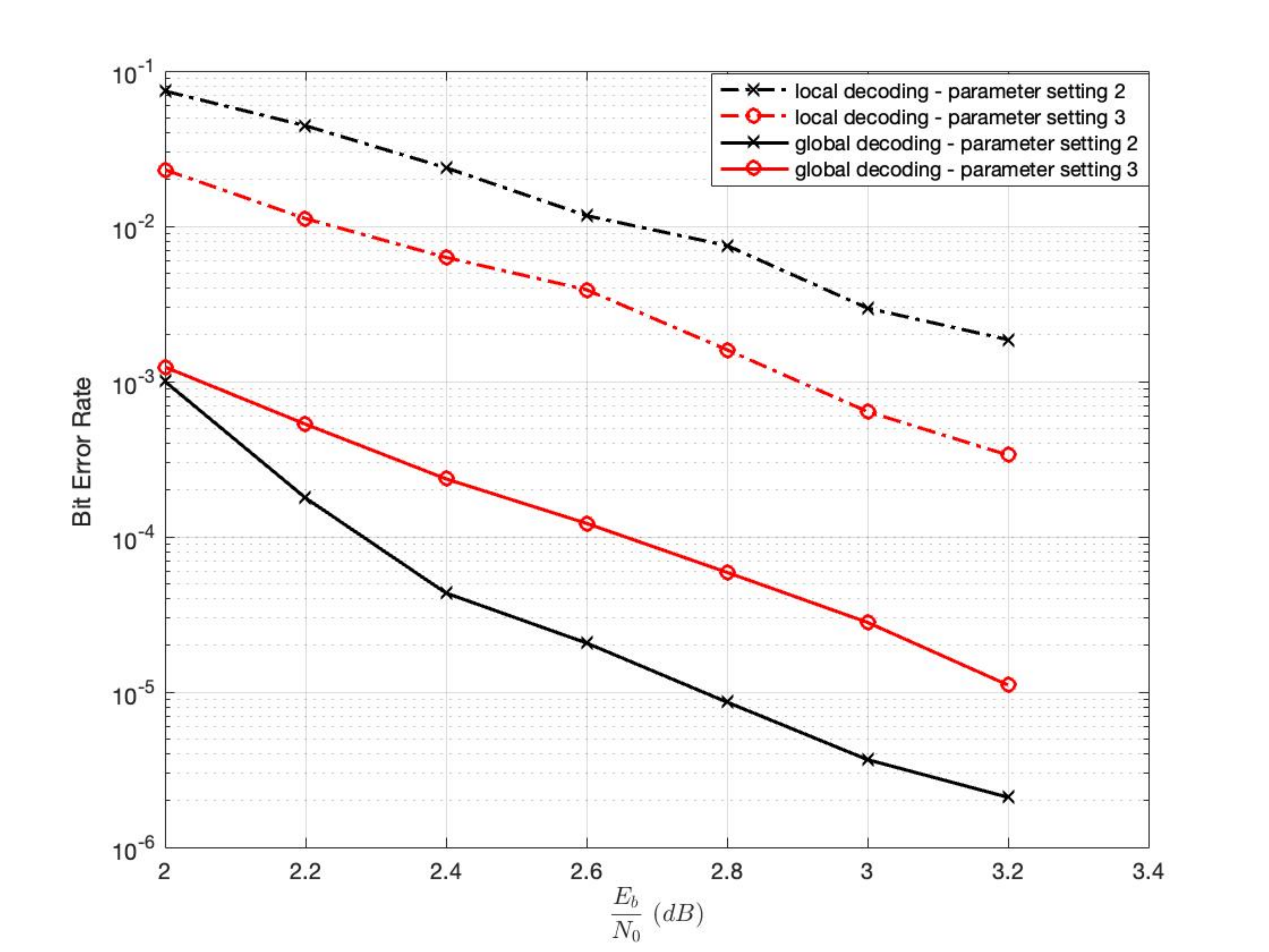}}
\caption{Trade-off between local and global decoding.}
\label{local global trade off}
\end{figure}

Fig.~\ref{local global trade off} illustrates this trade-off by comparing the BER performance of local-global decoding of 2-subblock constructions using parameter settings 2 and 3 in 
Table~\uppercase\expandafter{\romannumeral1}. Setting 2 uses an outer code of rate $R_{outer}{=}0.5625$ and length length $N_0{=}256$, whereas setting 3 uses an outer code with  rate $R_{outer}{=}0.53125$ and length $N_0{=}128$. As seen in Fig.~\ref{local global trade off}, using the outer code of lower rate and smaller length in setting~3 improves the local decoding performance, but worsens the global decoding performance, as compared to setting~2.  


\section{Conclusion}
 We proposed a concatenated architecture for local-global decoding of polar codes.
 The construction allows independent decoding of subblocks corresponding to inner polar codes, with global decoding facilitated by use of a  systematic  outer polar code. BER and FER simulation results for local-global decoding of 2 subblocks and 4 subblock show that global decoding provides performance comparable to that of a conventional polar code of the same overall rate and length under BP decoding, but with a penalty in the local decoding of the subblocks compared to a conventional polar code of the same rate and length. The trade-off between local and global decoding performance that can be achieved by adjusting system parameters was illustrated. These results demonstrate that the local-decoding paradigm proposed for graph-based codes~\cite{RamCas2018}, \cite{RamCas_SCLDPC} can be extended to polar codes. Several design optimization scenarios remain to be explored.

\section*{Acknowledgment}
The authors would like to thank Samsung Electronics Co., Ltd. for supporting this work.



\end{document}